\documentclass[floatfix,amssymb,prl,twocolumn,superscriptaddress,nofootinbib]{revtex4-1}
		
\usepackage{amssymb,amsmath,verbatim,mathtools,needspace,enumitem,etoolbox,graphicx,physics,microtype,afterpage,bm,soul}
\usepackage[dvipsnames]{xcolor}
\definecolor{linkcolor}{rgb}{0.0,0.3,0.5}
\usepackage[unicode, colorlinks=true, linkcolor=linkcolor, citecolor=linkcolor, filecolor=linkcolor,urlcolor=linkcolor, pdfusetitle]{hyperref}
\usepackage[all]{hypcap}
\usepackage[T1]{fontenc}
\usepackage[utf8]{inputenc}
\usepackage{tabularx}
\usepackage{float}
\usepackage[switch]{lineno} 
\usepackage[normalem]{ulem}
\usepackage{lmodern}
\allowdisplaybreaks
\usepackage{tikz}
\usepackage{color}
\usepackage{framed}
\usepackage{hyperref}
\hypersetup{colorlinks, citecolor=bluscuro, linkcolor=black, urlcolor=bluscuro}
\definecolor{rossos}{cmyk}{0,1,1,0.55}
\definecolor{bluscuro}{rgb}{0.15, 0.2, .85}
\definecolor{bluchiaro}{cmyk}{1,.3,0.,0.1}
\definecolor{ForestGreen}{rgb}{0.13, 0.55, 0.13}

\newcommand{\be}{\begin{equation}}
\newcommand{\ee}{\end{equation}}

\newcommand{\healpix}[0]{\texttt{HEALPix}}

\def\lsim{\mathrel{\rlap{\lower4pt\hbox{\hskip0.5pt$\sim$}}
    \raise1pt\hbox{$<$}}}         
\def\gsim{\mathrel{\rlap{\lower4pt\hbox{\hskip0.5pt$\sim$}}
    \raise1pt\hbox{$>$}}}         

\usepackage{etoolbox}
\usepackage{titlesec}

\titlespacing{\section}{2pt}{*.4}{*.4}
\titlespacing{\subsection}{0pt}{*.4}{*0.4}
\titlespacing{\subsubsection}{0pt}{*0.4}{*0.4}

\titlespacing{\section}{2pt}{*0.7}{*0.7}
\titlespacing{\subsection}{0pt}{*0.4}{*0.4}
\titlespacing{\subsubsection}{0pt}{*0.4}{*0.4}

\makeatletter
\providecommand\jnl@style{\rmfamily} 
\newcommand\ref@jnl[1]{{\jnl@style#1}}

\newcommand\aj{\ref@jnl{AJ}}
\newcommand\actaa{\ref@jnl{Acta Astron.}}
\newcommand\araa{\ref@jnl{ARA\&A}}
\newcommand\apjl{\ref@jnl{ApJ}}
\newcommand\apjs{\ref@jnl{ApJS}}
\newcommand\apss{\ref@jnl{Ap\&SS}}
\newcommand\aap{\ref@jnl{A\&A}}
\newcommand\aapr{\ref@jnl{A\&A~Rev.}}
\newcommand\aaps{\ref@jnl{A\&AS}}
\newcommand\azh{\ref@jnl{AZh}}
\newcommand\baas{\ref@jnl{BAAS}}
\newcommand\bac{\ref@jnl{Bull. astr. Inst. Czechosl.}}
\newcommand\caa{\ref@jnl{Chinese Astron. Astrophys.}}
\newcommand\cjaa{\ref@jnl{Chinese J. Astron. Astrophys.}}
\newcommand\icarus{\ref@jnl{Icarus}}
\newcommand\jcap{\ref@jnl{J. Cosmology Astropart. Phys.}}
\newcommand\jrasc{\ref@jnl{JRASC}}
\newcommand\memras{\ref@jnl{MmRAS}}
\newcommand\mnras{\ref@jnl{MNRAS}}
\newcommand\na{\ref@jnl{New A}}
\newcommand\nar{\ref@jnl{New A Rev.}}
\newcommand\pasa{\ref@jnl{PASA}}
\newcommand\pasp{\ref@jnl{PASP}}
\newcommand\pasj{\ref@jnl{PASJ}}
\newcommand\rmxaa{\ref@jnl{Rev. Mexicana Astron. Astrofis.}}
\newcommand\qjras{\ref@jnl{QJRAS}}
\newcommand\skytel{\ref@jnl{S\&T}}
\newcommand\solphys{\ref@jnl{Sol.~Phys.}}
\newcommand\sovast{\ref@jnl{Soviet~Ast.}}
\newcommand\ssr{\ref@jnl{Space~Sci.~Rev.}}
\newcommand\zap{\ref@jnl{ZAp}}
\newcommand\iaucirc{\ref@jnl{IAU~Circ.}}
\newcommand\aplett{\ref@jnl{Astrophys.~Lett.}}
\newcommand\apspr{\ref@jnl{Astrophys.~Space~Phys.~Res.}}
\newcommand\bain{\ref@jnl{Bull.~Astron.~Inst.~Netherlands}}
\newcommand\fcp{\ref@jnl{Fund.~Cosmic~Phys.}}
\newcommand\gca{\ref@jnl{Geochim.~Cosmochim.~Acta}}
\newcommand\grl{\ref@jnl{Geophys.~Res.~Lett.}}
\newcommand\jgr{\ref@jnl{J.~Geophys.~Res.}}
\newcommand\jqsrt{\ref@jnl{J.~Quant.~Spec.~Radiat.~Transf.}}
\newcommand\memsai{\ref@jnl{Mem.~Soc.~Astron.~Italiana}}
\newcommand\nphysa{\ref@jnl{Nucl.~Phys.~A}}
\newcommand\physrep{\ref@jnl{Phys.~Rep.}}
\newcommand\physscr{\ref@jnl{Phys.~Scr}}
\newcommand\planss{\ref@jnl{Planet.~Space~Sci.}}
\newcommand\procspie{\ref@jnl{Proc.~SPIE}}
\makeatother

\begin{document}

\title[DECADE+DES Y3 weak lensing mass map]{DECADE+DES Y3 Weak Lensing Mass Map: A 13,000 deg\(^2\) View of Cosmic Structure from 270 Million Galaxies}

\author{
M.~Gatti$^{1,*}$,
D.~Anbajagane$^{1,2,3}$,
C.~Chang$^{1,2,3}$,
D.~J.~Bacon$^{4}$,
J.~Prat$^{5,6}$,
M.~Adamow$^{7,8}$,
A.~Alarcon$^{9}$,
M.~R.~Becker$^{10}$,
J.~A.~Carballo\textendash Bello$^{11}$,
N.~Chicoine$^{3,12}$,
C.~Doux$^{13}$,
A.~Drlica\textendash Wagner$^{1,2,3,14}$,
P.~S.~Ferguson$^{15}$,
D.~Gruen$^{16,17}$,
R.~A.~Gruendl$^{7,8}$,
K.~Herron$^{18}$,
N.~Jeffrey$^{19}$,
D.~J.~James$^{20,21}$,
A.~Kov\'acs$^{22,23}$,
C.~E.~Mart\'inez\textendash V\'azquez$^{24}$,
P.~Massana$^{25}$,
S.~Mau$^{26,27}$,
J.~McCullough$^{28}$,
G.~E.~Medina$^{29,30}$,
B.~Mutlu\textendash Pakdil$^{18}$,
N.~E.~D.~No\"el$^{31}$,
A.~B.~Pace$^{32}$,
G.~Pollina$^{16}$,
A.~H.~Riley$^{33,34}$,
D.~J.~Sand$^{35}$,
L.~F.~Secco$^{1}$,
G.~S.~Stringfellow$^{36}$,
D.~Suson$^{37}$,
C.~Y.~Tan$^{38,1}$,
R.~Teixeira$^{3,39}$,
E.~J.~Tollerud$^{40}$,
M.~A.~Troxel$^{39}$,
L.~Whiteway$^{19}$,
A.~Zenteno$^{43}$,
Z.~Zhang$^{3,25,40}$
\\
{\textit{Affiliations are reported in the supplementary material}}
}


\begin{abstract}
We present the largest {galaxy} weak lensing mass map of the late-time Universe, reconstructed from 270 million galaxies in the DECADE and DES Year 3 datasets, covering 13,000 square degrees. We validate the map through systematic tests against observational conditions {(depth, seeing, etc.)}, finding {the map is statistically consistent with no contamination}. The large area covered by the mass map makes it a well-suited tool for cosmological analyses, cross-correlation studies and the identification of large-scale structure features. We demonstrate its potential by detecting cosmic filaments directly from the mass map for the first time and validating them through their association with galaxy clusters selected using the Sunyaev-Zeldovich effect from \textit{Planck} and ACT DR6.
\end{abstract}

\maketitle

\section{INTRODUCTION}
\label{sec:intro}

Weak gravitational lensing is one of the primary cosmological probes used by recent galaxy surveys \citep{Bartelmann2001,Mandelbaum2018}. By measuring the small distortions in galaxy shapes caused by the large-scale matter distribution between the observed galaxies and us, we can reconstruct the projected mass distribution, known as the convergence field or weak lensing mass map. Surveys such as the Kilo-Degree Survey (KiDS; \citep{Kuijken2015}), Hyper Suprime-Cam Subaru Strategic Program (HSC-SSP; \citep{Aihara2018}), Dark Energy Survey (DES; \citep{DES2016}), and {Dark Energy Camera All Data Everywhere} project (DECADE; \citep{Anbajagane2025a,Anbajagane2025b,Anbajagane2025c,Anbajagane2025d}) provide wide-area datasets enabling the reconstruction of these fields over large sky regions. 

Mapping the mass distribution through weak lensing is valuable for constraining the matter content of the Universe ($\Omega_{m}$), the amplitude of matter fluctuations ($\sigma_{8}$), and the nature of dark energy. Beyond traditional two-point statistics, which are widely used to constrain cosmology, mass maps encode additional information through non-Gaussian structures. This has motivated the development of alternative methods—including higher-order moments \citep{moments2021,Porth2021,G24,G25}, peak counts \citep{Harnois2024,Marques2024,Novaes2024,Jeffrey2025}, topological statistics \citep{Kratochvil2012,Grewal2022,Marques2024,Prat2025},
map-level compressions enabled by machine learning techniques 
\citep{Fluri2019,jeffrey_lfi,Jeffrey2025}—which extract information directly from mass maps. In addition to cosmology, mass maps offer a unique view of the large-scale structure, providing insights into the connection between galaxies, clusters, and the cosmic web, and serve as a key tool for cross-correlations with other structure tracers \citep{Hojjati2017,Davies2018,Miyazaki2018,Oguri2021,Kovacs2022,Kovacs2023}.

Various algorithms designed to construct mass maps exist, each with their own {advantages/disadvantages}. The Kaiser-Squires (KS; \citep{Kaiser1993}) method, widely used for mass mapping, provides a simple inversion of the shear field without assuming a prior on the convergence field. While computationally efficient, the lack of a prior leads to noise-dominated maps—{this is} acceptable in analyses where noise is forward-modeled, but limiting for detecting structures like voids and clusters. The Wiener filter method \citep{Wiener1949,Jeffrey2021} assumes a Gaussian prior on the convergence field, {which is} a reasonable large-scale approximation. Its \textit{maximum a posteriori} (MAP) estimate typically correlates better with the true mass map and yields posterior samples \citep[e.g.,][]{Kovacs2023} for uncertainty propagation. More advanced methods—using log-normal \citep{Fiedorowicz2022,Boruah2024}, wavelet-based \citep{Lanusse2016,Price2021,Starck2021}, or simulation-based  \citep{Remy2020,Jeffrey2020,Whitney2024,Boruah2025} priors—further improve reconstruction and often provide uncertainty samples, but at higher computational cost, limited resolution, or via planar approximations that are unsuitable for modern wide-area surveys requiring spherical methods. For this work, we therefore adopt the KS and Wiener approaches as representative baselines.

In this letter we present the largest galaxy weak-lensing mass map to date, covering $\sim$13{,}000 deg$^2$ from combining DECADE and DES Y3 data. Because both rely on DECam imaging \citep{Flaugher2015} and closely aligned pipelines, it is possible to coherently combine their weak lensing data. The resulting DECADE+DES map is $\sim3\times$ larger than prior efforts (DES Y3 $\sim$4{,}143 deg$^2$ \citep{Jeffrey_mass_map}; KiDS $\sim$1{,}347 deg$^2$\citep{Wright2025}; HSC $\sim$416 deg$^2$ \citep{Li2022}) while maintaining the uniformity required for structure finding and cosmology. The area gain reflects adding DECADE data to the DES Y3 footprint. In the following, we present our data, followed by the methodology used for map construction. We then introduce the DECADE+DES Y3 mass map, along with a series of null tests with systematic maps and a scientific application of the map, where we identify filaments.


\section{DATA}

\begin{figure*}
    \centering
    \includegraphics[width = 0.9\textwidth]{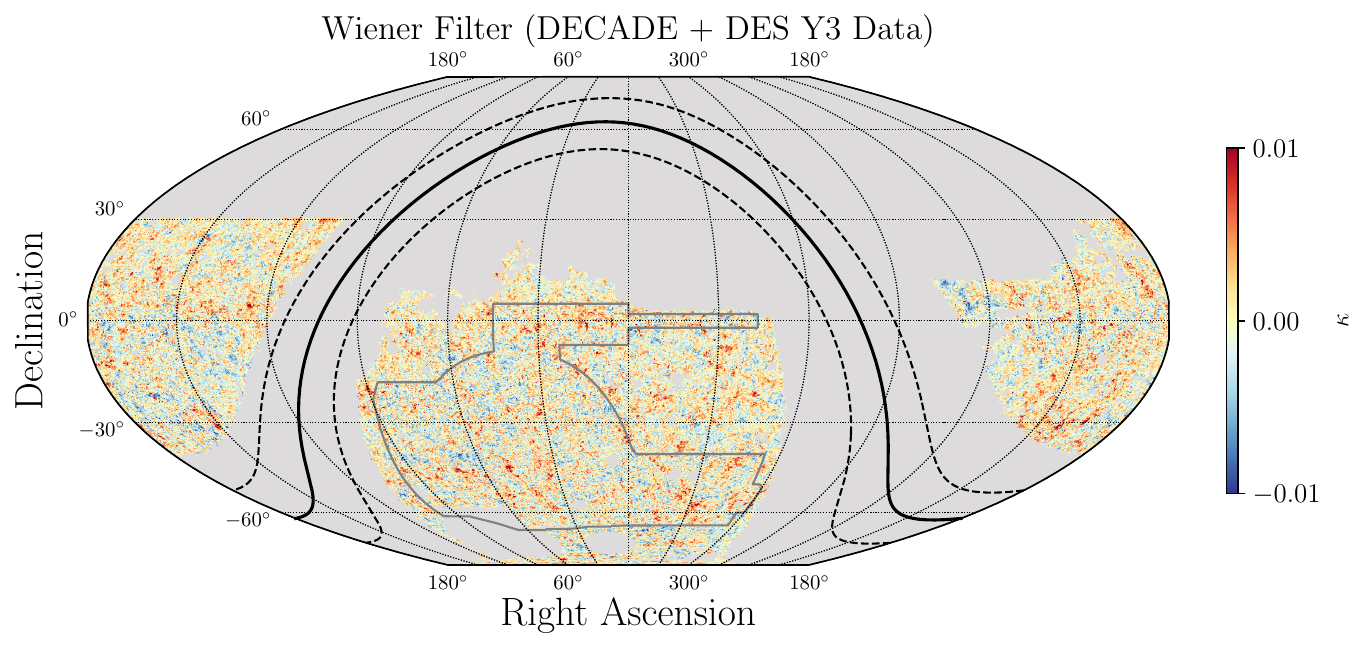}
    \caption{Wiener-MAP solution for the DECADE+DES Y3 mass map. Overdense lines of sight appear red; blue indicates underdense regions. The tank-shaped grey area marks the DES footprint; data outside come from DECADE.}
    \label{data_maps}
\end{figure*}

We use weak lensing shear catalogs from the Dark Energy Survey Year 3 \citep{Gatti2021} and the DECADE project \citep{Anbajagane2025d,Anbajagane202513k}, containing approximately 100 million and 170 million galaxies, respectively, covering a combined area of ~13,000 square degrees. {Both are based on DECam imaging. DES Y3 was designed as a dedicated survey, producing a more homogeneous dataset, whereas DECADE combines public DECam imaging mainly from structured wide-area programs (DELVE, DECaLS, DeROSITAS) designed for near-uniform coverage, along with smaller programs, resulting in greater variation in exposure time and image quality
\citep{Anbajagane2025a,Tan2025}}. In both cases, shear components ($\gamma_{1,2}$) are estimated using the \texttt{Metacalibration} method \citep{Sheldon2017,HuffMcal2017}, which self-calibrates shear measurements by correcting estimator response and selection effects.  While the catalogs were originally divided into four tomographic bins using self-organizing maps \citep{y3-sompz}, we use the full, unbinned samples to construct weak lensing mass map.

The catalogs are used to create shear maps, i.e., pixelized maps of the two shear components, using a \healpix{} pixelization \citep{GORSKI2005} with \texttt{NSIDE} $ = 1024$, roughly corresponding to $3.44~\mathrm{arcmin}$ pixels. The shear value in each pixel is computed as:
\begin{equation}
\label{eq:pixelvalue}
\gamma_{\rm obs}^{\nu} = \frac{\sum_{j=1}^{n}\epsilon_j^{\nu}w_{j}}{\bar{R}\sum_{j=1}^{n}w_j}, \,\, \nu=1,2,
\end{equation}
Here, $\nu$ denotes the two shear field components, $n$ is the total number of galaxies in the sample, $w_j$ is the per-galaxy inverse-variance weight \citep{Gatti2021}, and $\bar{R}$ is the mean \texttt{Metacalibration} response of the sample. We correct the DES and DECADE samples separately and use Eq.~\ref{eq:pixelvalue} to create shear maps which, though non-overlapping, connect with minimal discontinuity. The combined shear map is the starting point for map-making.  Adding DECADE reduces edge effects in the DES region of the final mass map relative to the fiducial DES Y3 map \citep{Jeffrey2021}.
While the response corrects most shear biases, a residual percent-level multiplicative bias—mainly from blending and usually estimated with image simulations \citep{Anbajagane2025a,y3-imagesims}—is not corrected in our maps. This standard practice leaves the correction to users; its small size does not affect our conclusions. Any non-zero mean shear is removed from the catalog prior to map-making.

\section{MASS MAP INFERENCE}\label{sect:map}

Weak lensing mass mapping aims to reconstruct the convergence field \(\kappa\), which describes the projected mass distribution, from noisy shear measurements \(\gamma_{\text{obs}}\). This is commonly formulated as a Bayesian inference problem \citep{Jeffrey_mass_map,Remy2023}:
\[
p(\kappa | \gamma_{\text{obs}}) \propto p(\gamma_{\text{obs}} | \kappa) \, p(\kappa),
\]
where  \(p(\kappa)\) is a prior on the convergence field, and \(p(\gamma_{\text{obs}} | \kappa)\) is the likelihood.

{A common approximation for the \healpix{} map pixel likelihood is to assume it is multi-variate Gaussian\footnote{This holds in the noise-dominated limit with many galaxies per pixel, as expected from the Central Limit Theorem. Most mass-mapping methods adopt this assumption; our maps contain on average $\sim$60 galaxies per pixel.}}:
\begin{equation} \label{eq:likelihood}
\log p( {\gamma_{\mathrm{obs}}} | {\kappa} ) = - \frac{1}{2} ( {\gamma_{\mathrm{obs}}} - \mathbf{A} {\kappa} )^\dagger \mathbf{N}^{-1} ( {\gamma_{\mathrm{obs}}} - \mathbf{A} {\kappa} )  + \text{const.}
\end{equation}
The noise covariance matrix \(\mathbf{N}\) is assumed to be dominated by shape noise and to be diagonal (i.e., pixel noise is uncorrelated). Masked pixels are assigned infinite variance in \(\mathbf{N}\), ensuring they do not contribute to the likelihood. In the above, we assumed a linear data model:
\begin{equation}\label{eq:linearmod}
{\gamma_{\mathrm{obs}}} = \mathbf{A} {\kappa} + \mathbf{n},
\end{equation}
with $\mathbf{n}$ denoting the pixel-level noise and $\mathbf{A}$ representing the linear transformation from the true (noise-free and full-sky) convergence field to the shear field \citep{Chang2018}. While Eq. \ref{eq:linearmod} is valid in both real and harmonic space, in harmonic space this transformation can be written as
\begin{equation}
\label{eq:mass_map_operator}
\hat{\gamma}_{\ell m} = -\sqrt{\frac{(\ell - 1)(\ell + 2)}{\ell(\ell + 1)}} \, \hat{\kappa}_{\ell m},
\end{equation}
 where \(\hat{\gamma}_{\ell m}\) and \(\hat{\kappa}_{\ell m}\) are the harmonic coefficients of the shear and convergence fields, respectively. The harmonic coefficients are typically decomposed into real and imaginary parts, as \(\hat{\kappa}_{\ell m} = \hat{\kappa}_{E,\ell m} + i \hat{\kappa}_{B,\ell m}\) and \(\hat{\gamma}_{\ell m} = \hat{\gamma}_{E,\ell m} + i \hat{\gamma}_{B,\ell m}\).   To first order in lensing and in the absence of noise, the imaginary components (B-modes) are expected to vanish. {However, in the presence of masking, the conversion from observed shear to convergence can induce leakage from E-modes into B-modes} \citep{Jeffrey2021}.
Here we consider two mass-mapping methods:

\noindent  \textbf{Kaiser-Squires (KS)}: the KS inversion corresponds to the MAP solution for a flat prior \(p(\kappa) \propto 1\). The practical implementation of the KS method on the sphere involves decomposing the observed spin-2 shear field \(\gamma\) into a curl-free E-mode component and a divergence-free B-mode component. {The convergence field is then recovered by applying Eq. \ref{eq:mass_map_operator}, as described in the previous section, to obtain the harmonic coefficients $\hat{\kappa}_{\mathrm{E}, \ell m}$ and $\hat{\kappa}_{\mathrm{B}, \ell m}$. A spin-0 spherical harmonic transform is then applied separately to $\hat{\kappa}_{\mathrm{E}, \ell m}$ and $\hat{\kappa}_{\mathrm{B}, \ell m}$ to obtain their real-space counterparts, $\kappa_{\mathrm{E}}(\theta, \varphi)$ and $\kappa_{\mathrm{B}}(\theta, \varphi)$, respectively.} KS is typically followed by smoothing, effectively imposing a non-flat prior on \( p(\kappa) \) determined by the smoothing type. Smoothing serves different purposes by application. For mass-map inference, it boosts signal-to-noise (S/N) by suppressing small-scale noise, yielding maps that better track the true convergence and aiding structure identification. For cross-correlations or simulation-based inference, it enforces scale separation; multiple kernels (Gaussian, top-hat, wavelet) are often applied and retained. The optimal smoothing scale is application-dependent and usually determined via simulations \citep{Jeffrey_mass_map}.

\noindent  \textbf{Wiener filter}: the Wiener filter \citep{Wiener1949,zaroubi_wiener,Jeffrey2021} assumes that the convergence field follows a Gaussian random field prior and has no B-mode power. The Wiener MAP solution is given by:
\begin{equation}
\hat{\kappa}_W = \mathbf{S}_\kappa \mathbf{A}^\dagger \left( \mathbf{A} \mathbf{S}_\kappa \mathbf{A}^\dagger + \mathbf{N} \right)^{-1} \gamma_{\text{obs}},
\end{equation}
where $\mathbf{S}_\kappa$ is the signal covariance matrix.
The Wiener filter is implemented through the code 
\texttt{Dante} \citep{Ramanah2019}, using a class of methods based on messenger fields \citep{jasche2015matrix, alsing2017cosmological, jeffrey_heavens_fortio}, which iteratively transform between pixel space—where the noise covariance matrix \(\mathbf{N}\) is diagonal—and harmonic space—where the signal covariance matrix \(\mathbf{S}_\kappa\) is diagonal (i.e., it corresponds to the power spectrum of the convergence field). Additionally, \texttt{Dante} allows us to draw samples from the posterior \( p({\kappa} | {\gamma_{\mathrm{obs}}}) \) through constrained realizations. These samples represent possible realizations of \(\kappa\) that are consistent with both the observed data and the Gaussian prior assumed in the Wiener filter approach.

{We validate the KS and Wiener filter methods on mock DES+DECADE weak lensing mass map, as detailed in the supplementary material, by performing standard tests from the literature and evaluating how well the methods recover the true underlying convergence field.}

\section{RESULTS ON DATA}

\begin{figure}
    \centering

    \includegraphics[width =  0.9\columnwidth]{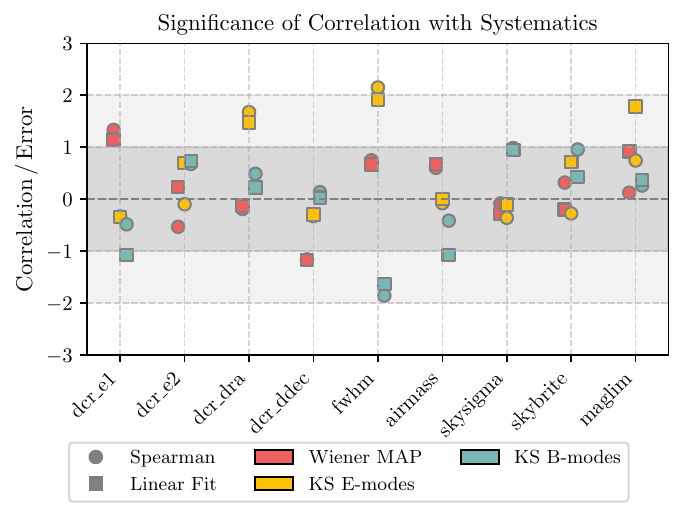}
    \caption{Significance of the test {for linear dependence (squares) and Spearman rank correlation (circles) with systematics} for the Wiener MAP solution and the KS E-mode and B-mode maps. From left to right, the systematic maps considered are: differential chromatic refraction (DCR) in the \(e_1\) and \(e_2\) directions (with respect to the focal plane orientation), DCR in right ascension (R.A.) and declination (Dec.), the seeing (PSF FWHM), airmass, the standard deviation of the sky background, the sky brightness, and the magnitude limit (or survey depth). For this test we removed modes with $\ell$<10, which are dominated by survey geometry and contain no useful cosmological information.}
    \label{syst}
\end{figure}

 We present the completed DECADE+DES Y3 weak lensing mass map in Fig.~\ref{data_maps}. To produce the map, we first constructed the DECADE and DES Y3 shear maps, added them, and then we obtained both the KS map and the Wiener filter MAP solution and samples following the previous section. For the Wiener prior, we assume the FLAGSHIP cosmology \citep{Flagship},\footnote{$w=-1$, $\Omega_{\rm m} = 0.319$, $A_{\rm s}= 2.1 \, 10^{-9}$ (corresponding to $\sigma_8 \sim 0.813$), $\Omega_{\rm b} = 0.049$, $n_{\rm s} = 0.96 $, 
$h = 0.67$, $\Omega_{\Lambda} = 0.681 - \Omega_{\rm r} - \Omega_{\nu}$ , with a radiation density $\Omega_{\rm r} = 5.509 \times 10^{-5}$, and a contribution from massive neutrinos $\Omega_{\rm \nu} = 0.00140343$} consistent with constraints from the DECADE and DES Y3 cosmic shear analyses \citep{Anbajagane2025d,y3-cosmicshear1,y3-cosmicshear2}. The DES Y3 mass map, located at the center, transitions smoothly into the DECADE map, without any visible discontinuity between the two surveys.

\subsection{Systematic tests on data maps}

We test for spurious correlations between our mass map and quantities that are not expected to correlate with convergence, such as observing conditions. These tests are particularly important for a dataset like DECADE, which is obtained as a combination of different programs. While many null tests have already been performed on the DECADE shear catalog \citep{Anbajagane_shearcat}, the DES Y3 catalog \citep{Gatti2021}, and the DES Y3 mass map \citep{Jeffrey_mass_map}, we carry out additional tests on the combined weak lensing mass map.

To test for residual contamination in our mass map, we evaluate correlations with various observing conditions, using two complementary statistical tests: a linear fit and the Spearman rank correlation. The linear fit quantifies how well the mass map \(\kappa\) can be modeled as a linear function of a systematic map \(S\), i.e.,
\begin{equation}
\kappa = a + b \cdot S,
\end{equation}
where \(b\) is the slope of the best-fit line. A statistically significant deviation of \(b\) from zero indicates a potential additive systematic trend.

The Spearman rank correlation coefficient \(\rho\) measures the degree of monotonic relationship between \(\kappa\) and \(S\), based on the ranked values of the two maps. The rank of a value is its position in the sorted list of all unmasked pixels—for example, the pixel with the lowest value gets rank 1, the next lowest gets rank 2, and so on.

The Spearman coefficient is then given by:
\begin{equation}
\rho_{\rm S} = 1 - \frac{6 \sum_{i=1}^{N} d_i^2}{N(N^2 - 1)},
\end{equation}
where \(d_i\) is the difference between the ranks of \(\kappa_i\) and \(S_i\), and \(N\) is the number of unmasked pixels. Unlike the linear fit, this test is sensitive to both linear and non-linear monotonic relationships, such as logarithmic or saturation-like trends. The Spearman test is therefore a generalization of the Pearson correlation, capturing a broader class of dependencies while being more robust to outliers and non-Gaussian distributions.

Both tests are designed to detect additive systematics in the reconstructed convergence maps but are insensitive to multiplicative biases. We apply them to the Wiener-filtered MAP reconstruction and the KS maps, including both E- and B-modes. Fig.~\ref{syst} shows the significance of the linear slope (\(b / \sigma_{\rm b}\)) and Spearman rank correlation (\(\rho_{\rm S}/\sigma_{\rho}\)) with systematic maps. For KS maps, we show results smoothed at 20 arcminutes—the scale expected to best match the true convergence field \citep{Jeffrey_mass_map}—but we also explored smoothing scales from the pixel size up to 200 arcminutes, consistent with typical scales used in the cosmological shear analyses. When computing $p$-values for correlations with all systematics, we find no significant bias, with $p$-values exceeding $p>0.01$, based on both Fig.~\ref{syst} and the additional smoothed KS maps. Error bars for both tests are estimated from 100 jackknife regions, accounting for spatial correlations and sample variance. 

\subsection{Structures in the reconstructed maps}
\begin{figure*}
    \centering
    \includegraphics[width =  0.9\textwidth]{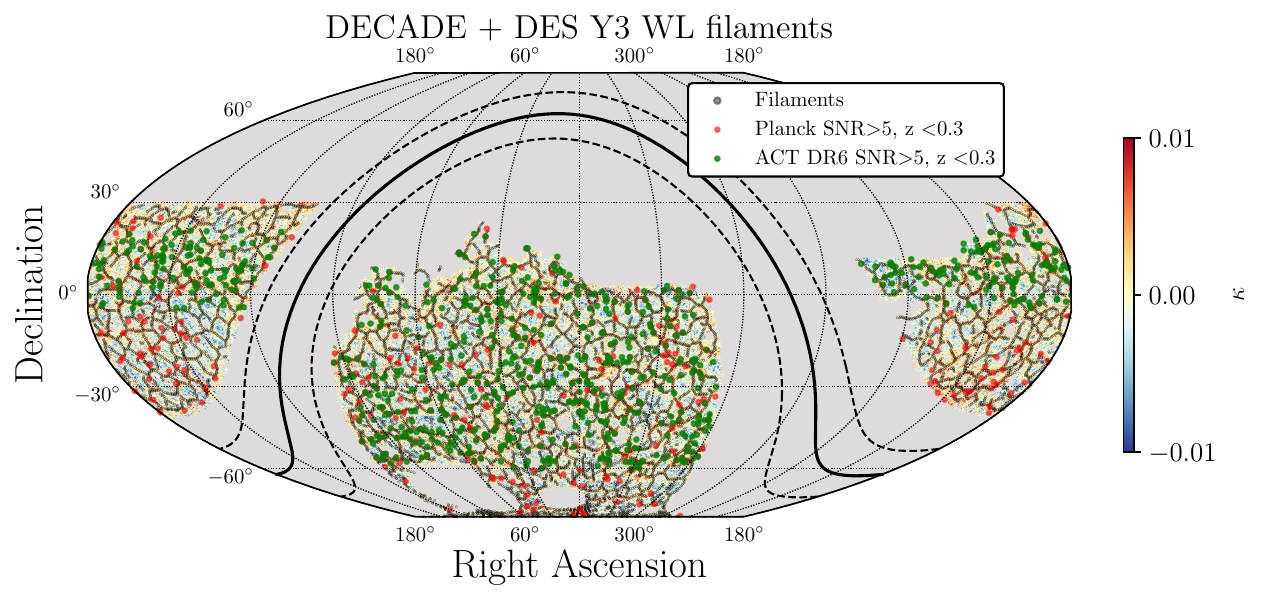}
    \caption{Filaments identified from the Wiener MAP, with SZ-selected clusters from \textit{Planck} and ACT DR6 overlaid.}
    \label{filaments2}
\end{figure*}


\begin{figure}
    \centering
    \includegraphics[width =  0.9\columnwidth]{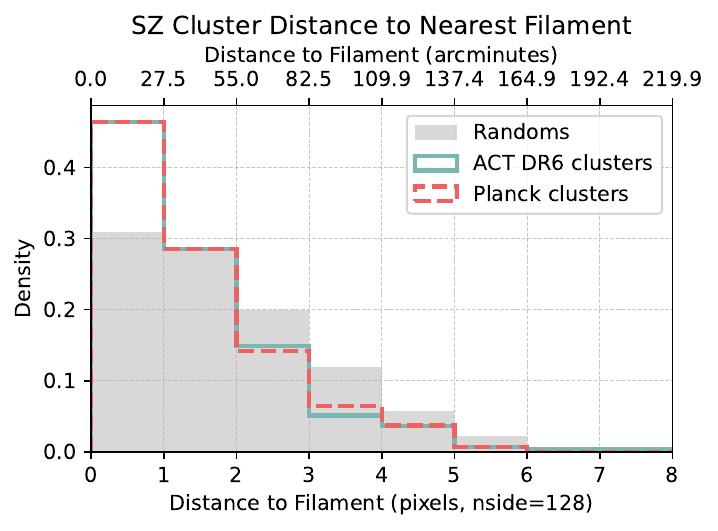}
    \caption{Distribution of the distance to the nearest filament pixel for SZ-selected clusters from \textit{Planck} and ACT DR6. The distribution is compared to that of ACT DR6 randoms. }
    \label{filaments_stats2}
\end{figure}

As an example usage, we use the combined weak lensing map to detect filaments. Weak lensing maps have previously been used to detect clusters by identifying high S/N peaks \citep{Miyazaki2018,Oguri2021}, and to select line-of-sight projected voids \citep{Davies2018} as an alternative to spectroscopic or photometric galaxy samples, showing certain advantages over the latter. However, weak lensing mass maps have never been used to identify filamentary structures, for which spectroscopic (or photometric) galaxy samples have typically been preferred \citep{Malavasi2020,Laigel2025}. Similar to voids, filament-finder algorithms can be applied directly to weak-lensing mass maps, avoiding the need to model galaxy bias and selection, though at the cost of increased sensitivity to shape noise, masking, and line-of-sight projection.

To identify filaments, we use the \texttt{SCONCE} algorithm \citep{Zhang2022}, a cosmic web finder designed for spherical and conical geometries. \texttt{SCONCE} generalizes the SCMS method \citep{Moews2021,Carron2022}, widely used to trace filaments as density ridges, and improves recovery of structures on the sphere, especially at high declinations. Originally developed for galaxy samples, it estimates the density field and identifies filaments via adaptive gradient ascent. For weak-lensing maps, we evaluate the map at each pixel center and provide \texttt{SCONCE} with pixel coordinates weighted by map values. The method requires a smoothing scale, below which variations are ignored. Following the scale maximizing the Pearson correlation between KS and true convergence for DES/DECADE noise ($\sim$20–30 arcmin; \citep{Jeffrey2021}), we downgrade the map to \texttt{NSIDE}=128 (pixel size $\sim$28 arcmin) before sampling.

Fig. \ref{filaments2} shows the filamentary structure identified by \texttt{SCONCE}.  By construction, the filaments avoid the most underdense regions of the map, instead tracing overdense ridges. Masking can affect the detection of filaments near survey boundaries, although a systematic assessment of this impact is left for future work. The strength of the signal can be quantified by comparing the average S/N of pixels belonging to filaments with that of the remaining pixels: on average, filament pixels have an S/N about 0.5 higher than the rest of the map (the map itself has a S/N$\sim$0 with root-mean-square fluctuations of $\sim$1).  In Fig. \ref{filaments2}, we plot the filamentary structure detected in the DES+DECADE mass map with galaxy clusters detected via the Sunyaev-Zeldovich effect (SZ) by \textit{Planck} \citep{Planck_cluster2016} and ACT DR6 \citep{Hilton2021} overlaid. We consider clusters with S/N > 5, restricting the sample to $z < 0.3$ where the weak-lensing mass map is most sensitive.  

Visually, clusters align with filaments and their intersections. This trend is quantified in Fig.~\ref{filaments_stats2}, where we show the distribution of distances between clusters and the nearest filament pixel, compared to the distribution for a random sample. For both \textit{Planck} and ACT DR6 SZ-selected clusters, the distributions differ significantly from that of the random points, with most clusters lying within 1–2 pixels of the nearest filament. This supports the interpretation that the structures identified by \texttt{SCONCE} trace real projected filamentary structures associated with overdense regions where clusters preferentially reside. 
While detailed analysis is beyond this letter, these results suggest weak-lensing filaments can be used for cosmological applications—probing structure growth, cross-correlations studies, and stacking gas-sensitive observables (e.g., Compton-$y$ maps) aligned with large-scale structure—to test feedback models \citep{Lokken2024}.



\section{CONCLUSIONS}
\label{sec:summary}

In this letter, we presented the largest galaxy weak-lensing mass map to date, covering $\sim$13,000 deg² and obtained from 270 million DECADE+DES Y3 galaxies. Both datasets use DECam imaging and similar data processing pipelines, enabling a coherent combination of their weak lensing data. The map is reconstructed on the sphere with two methods: the Kaiser–Squires approach (with small-scale smoothing) and the Wiener filter, which also generates posterior samples through constrained realizations of the shear data. We validate the reconstructed map through systematic tests against observational conditions, finding no significant contamination. The map’s large area makes it well-suited for a wide range of applications, e.g., identifying large-scale structures, cross-correlations with other cosmological probes, and extracting non-Gaussian information beyond two-point statistics. Using this map, we identified filaments, a task typically done with spectroscopic or photometric galaxy samples, but never before with weak lensing maps. We showed that the recovered filaments have a higher S/N than the rest of the map, and that SZ-selected clusters from \textit{Planck} and ACT DR6 preferentially align with them and their intersections.

While future stage IV surveys such as the Vera C. Rubin Observatory's Legacy Survey of Space and Time (LSST) and the Euclid mission will eventually deliver mass maps with even higher precision and similar footprints, this combined DECADE+DES Y3 map already provides an unprecedented large weak lensing dataset. Its wide area and robust construction make it a valuable resource for current and upcoming analyses before stage IV datasets become available.
\textit{Data availability}: The mass map will be shared upon acceptance of this manuscript.

\bibliographystyle{JHEP}
\bibliography{bibliography}

\section{SUPPLEMENTARY MATERIAL: VALIDATION ON SIMULATIONS}\label{sec:appendix}

We validate the KS and Wiener methods using mock catalogs derived from a full-sky N-body simulation run with the PKDGRAV3 code \citep{Potter2016} at the FLAGSHIP cosmology \citep{Flagship}. The simulation adopts a similar setup to \citep{Flagship} but at lower resolution, with $1350^3$ particles in a 1250 $h^{-1}$ Mpc box and lens planes generated at $\sim100$ redshifts from $z=49$ to $z=0$, equally spaced in proper time. Shear mock catalogs for the DECADE and DES Y3 datasets are created following the pipeline of \citep{Gatti2024}: lens planes are converted into convergence planes under the Born approximation, then into shear planes via the full-sky KS algorithm. Redshift-averaged noiseless shear planes are obtained by weighting each shell according to the redshift distributions of DECADE and DES Y3. In this process, we generated the simulated DECADE and DES Y3 maps separately, due to their slightly different redshift distributions (mean redshift $\sim$0.613 for DECADE and $\sim$0.630 for DES). To generate noisy shear catalogs, we used the real DECADE and DES Y3 catalogs by randomly rotating galaxy ellipticities to erase the true shear, then adding the simulated shear at each galaxy's position according to its associated pixel. The original inverse-variance weights were preserved to compute the intrinsic ellipticity. This procedure yields a simulated DECADE+DES Y3 catalog with the same number density, shape noise, and weights as the real data.

\begin{table}
\[
\begin{array}{|c|c|c|}
\hline
\text{Method} & \text{RMSE} & P_\kappa \ (\text{Pearson Coefficient}) \\
\hline
\text{Kaiser-Squires KS} & 0.0371 & 0.193 \\
\text{Wiener} & 0.0073 & 0.411 \\
\text{KS + Smoothing } 20' & 0.0082 & 0.340 \\
\text{KS + Smoothing } 25' & 0.0079 & 0.337 \\
\hline
\end{array}
\]
\caption{Comparison of RMSE and Pearson coefficients between the recovered simulated maps and the true noiseless convergence map for the KS (with and without smoothing) and Wiener MAP solutions.}
\label{table_}
\end{table}

\begin{figure}
    \centering
    \includegraphics[width =  \columnwidth]{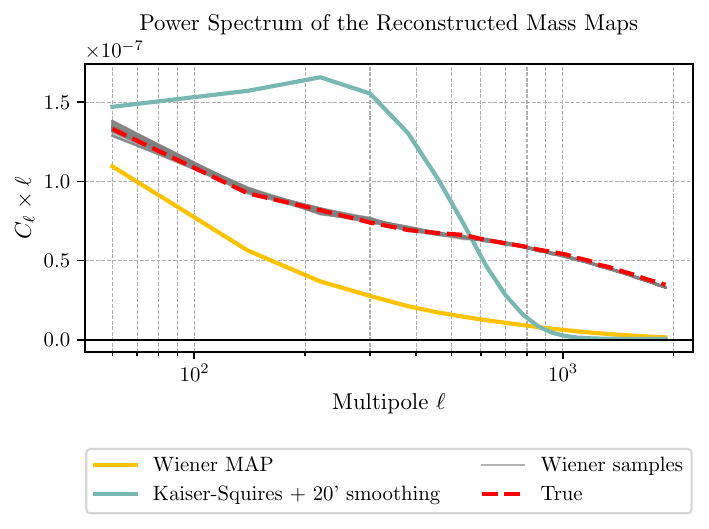}
    \caption{Power spectrum of the recovered simulated mass map, compared to the power spectrum of the true, noiseless convergence map.}
    \label{power_spectra}
\end{figure}

We characterise the KS and Wiener methods by comparing them against standard statistical measures using simulated maps, following \citep{Jeffrey_mass_map} and common practice in the literature. Specifically, we consider:

\begin{enumerate}
    \item The root mean square error (RMSE) between the true and reconstructed convergence fields.
    \item The Pearson correlation coefficient quantifying their correlation.
    \item The power spectrum of the reconstructed maps (and of the Wiener samples), compared to that of the true convergence field.
\end{enumerate}

We report the RMSE and Pearson coefficients in Table~\ref{table_}. Smoothing the KS map improves both metrics compared to the raw version, despite some loss of small-scale information. The Pearson coefficient is maximized at a smoothing scale of 20 arcminutes, while the RMSE reaches its minimum at 25 arcminutes. The Wiener MAP solution consistently outperforms the smoothed KS maps. 

The Wiener samples provide an estimate of the uncertainty in the Wiener MAP solution, i.e., its deviation from the true, noiseless convergence map. Although they assume a Gaussian convergence field, they approximate the uncertainty reasonably well. We verified that the RMS of the residuals matches the distribution from the Wiener samples within 10 percent, indicating that they provide a robust uncertainty estimate despite the non-Gaussianity of the true field.

We compare the recovered power spectra to the true noiseless convergence power spectrum in Fig.~\ref{power_spectra}. Neither the smoothed KS nor the Wiener MAP solution fully recovers the input spectrum, as their differences reflect the influence of priors and filtering on the \textit{maximum a posteriori} reconstruction. Gaussian smoothing in the KS map suppresses small-scale power while boosting large scales, whereas the Wiener MAP spectrum is suppressed across all scales. In both cases, this bias is expected and consistent with previous mass-mapping tests (e.g.\ DES Y3; \citep{Jeffrey2021}): MAP-based reconstructions yield the most probable map, but nonlinear statistics such as the power spectrum are not guaranteed to be unbiased. By contrast, averaging over Wiener posterior samples $p(\kappa|\gamma_{\mathrm{obs}})$ recovers the true spectrum by construction (although the unbiased recovery applies only to the power spectrum in the Wiener case). MAP reconstructions remain useful — for visualization or for cosmological inference if the prior’s impact is known and can be modeled.

\section{ACKNOWLEDGEMENTS}
The DELVE project is partially supported by Fermilab LDRD (L2019-011), the NASA Fermi Guest Investigator Program Cycle 9 No. 91201, and the National Science Foundation under Grant No. AST-2108168, AST-2108169, AST-2307126, and AST-2407526. This research award is partially funded by a generous gift of Charles Simonyi to the NSF Division of Astronomical Sciences. The award is made in recognition of significant contributions to Rubin Observatory’s Legacy Survey of Space and Time.

Funding for the DES Projects has been provided by the U.S. Department of Energy, the U.S. National Science Foundation, the Ministry of Science and Education of Spain, 
the Science and Technology Facilities Council of the United Kingdom, the Higher Education Funding Council for England, the National Center for Supercomputing 
Applications at the University of Illinois at Urbana-Champaign, the Kavli Institute of Cosmological Physics at the University of Chicago, 
the Center for Cosmology and Astro-Particle Physics at the Ohio State University,
the Mitchell Institute for Fundamental Physics and Astronomy at Texas A\&M University, Financiadora de Estudos e Projetos, 
Funda{\c c}{\~a}o Carlos Chagas Filho de Amparo {\`a} Pesquisa do Estado do Rio de Janeiro, Conselho Nacional de Desenvolvimento Cient{\'i}fico e Tecnol{\'o}gico and 
the Minist{\'e}rio da Ci{\^e}ncia, Tecnologia e Inova{\c c}{\~a}o, the Deutsche Forschungsgemeinschaft and the Collaborating Institutions in the Dark Energy Survey. 

The Collaborating Institutions are Argonne National Laboratory, the University of California at Santa Cruz, the University of Cambridge, Centro de Investigaciones Energ{\'e}ticas, 
Medioambientales y Tecnol{\'o}gicas-Madrid, the University of Chicago, University College London, the DES-Brazil Consortium, the University of Edinburgh, 
the Eidgen{\"o}ssische Technische Hochschule (ETH) Z{\"u}rich, 
Fermi National Accelerator Laboratory, the University of Illinois at Urbana-Champaign, the Institut de Ci{\`e}ncies de l'Espai (IEEC/CSIC), 
the Institut de F{\'i}sica d'Altes Energies, Lawrence Berkeley National Laboratory, the Ludwig-Maximilians Universit{\"a}t M{\"u}nchen and the associated Excellence Cluster Universe, 
the University of Michigan, NSF's NOIRLab, the University of Nottingham, The Ohio State University, the University of Pennsylvania, the University of Portsmouth, 
SLAC National Accelerator Laboratory, Stanford University, the University of Sussex, Texas A\&M University, and the OzDES Membership Consortium.

The DES data management system is supported by the National Science Foundation under Grant Numbers AST-1138766 and AST-1536171.
The DES participants from Spanish institutions are partially supported by MICINN under grants ESP2017-89838, PGC2018-094773, PGC2018-102021, SEV-2016-0588, SEV-2016-0597, and MDM-2015-0509, some of which include ERDF funds from the European Union. IFAE is partially funded by the CERCA program of the Generalitat de Catalunya.
Research leading to these results has received funding from the European Research
Council under the European Union's Seventh Framework Program (FP7/2007-2013) including ERC grant agreements 240672, 291329, and 306478.
We  acknowledge support from the Brazilian Instituto Nacional de Ci\^encia
e Tecnologia (INCT) do e-Universo (CNPq grant 465376/2014-2).

Fermilab is managed by FermiForward Discovery Group, LLC under Contract No. 89243024CSC000002 with the U.S. Department of Energy, Office of Science, Office of High Energy Physics. The United States Government retains and the publisher, by accepting the article for publication, acknowledges that the United States Government retains a non-exclusive, paid-up, irrevocable, world-wide license to publish or reproduce the published form of this manuscript, or allow others to do so, for United States Government purposes.

\section{AFFILIATIONS}
\noindent *marcogatti29@gmail.com\\
{$^{1}$ Kavli Institute for Cosmological Physics, University of Chicago, Chicago, IL 60637, USA}\\
{$^{2}$ NSF-Simons AI Institute for the Sky (SkAI), 172 E. Chestnut St., Chicago, IL 60611, USA}\\
{$^{3}$ Department of Astronomy and Astrophysics, University of Chicago, Chicago, IL 60637, USA}\\
{$^{4}$ Institute of Cosmology and Gravitation, University of Portsmouth, Burnaby Rd, Portsmouth, UK}\\
{$^{5}$ Nordita, KTH Royal Institute of Technology and Stockholm University, Hannes Alfv\'ens v\"ag 12, SE\textendash10691 Stockholm, Sweden}\\
{$^{6}$ Oskar Klein Centre, Department of Physics, Stockholm University, SE\textendash106 91 Stockholm, Sweden}\\
{$^{7}$ Center for Astrophysical Surveys, National Center for Supercomputing Applications, 1205 West Clark St., Urbana, IL 61801, USA}\\
{$^{8}$ Department of Astronomy, University of Illinois at Urbana\textendash Champaign, 1002 W. Green St., Urbana, IL 61801, USA}\\
{$^{9}$ Institute of Space Sciences (ICE, CSIC), Campus UAB, 08193 Barcelona, Spain}\\
{$^{10}$ Argonne National Laboratory, 9700 South Cass Ave., Lemont, IL 60439, USA}\\
{$^{11}$ Instituto de Alta Investigaci\'on, Universidad de Tarapac\'a, Casilla 7D, Arica, Chile}\\
{$^{12}$ Department of Physics and Astronomy, University of Pittsburgh, 3941 O’Hara St., Pittsburgh, PA 15260, USA}\\
{$^{13}$ Universit\'e Grenoble Alpes, CNRS, LPSC\textendash IN2P3, 38000 Grenoble, France}\\
{$^{14}$ Fermi National Accelerator Laboratory, P.O. Box 500, Batavia, IL 60510, USA}\\
{$^{15}$ DiRAC Institute, Department of Astronomy, University of Washington, Seattle, WA 98195, USA}\\
{$^{16}$ University Observatory, Faculty of Physics, Ludwig\textendash Maximilians\textendash Universit\"at M\"unchen, Scheinerstr. 1, 81679 Munich, Germany}\\
{$^{17}$ Excellence Cluster ORIGINS, Boltzmannstr. 2, 85748 Garching, Germany}\\
{$^{18}$ Department of Physics and Astronomy, Dartmouth College, Hanover, NH 03755, USA}\\
{$^{19}$ University College London, London, UK}\\
{$^{20}$ ASTRAVEO LLC, PO Box 1668, Gloucester, MA 01931, USA}\\
{$^{21}$ Applied Materials Inc., 35 Dory Road, Gloucester, MA 01930, USA}\\
{$^{22}$ MTA\textendash CSFK \emph{Lend\"ulet} “Momentum” Large\textendash Scale Structure Research Group, Konkoly Thege Mikl\'os \'ut 15\textendash17, H\textendash1121 Budapest, Hungary}\\
{$^{23}$ Konkoly Observatory, HUN\textendash REN Research Centre for Astronomy and Earth Sciences, Konkoly Thege Mikl\'os \'ut 15\textendash17, H\textendash1121 Budapest, Hungary}\\
{$^{24}$ NSF NOIRLab, 670 N. A‘ohoku Place, Hilo, Hawai‘i 96720, USA}\\
{$^{25}$ NSF NOIRLab, Casilla 603, La Serena, Chile}\\
{$^{26}$ Department of Physics, Stanford University, 382 Via Pueblo Mall, Stanford, CA 94305, USA}\\
{$^{27}$ Kavli Institute for Particle Astrophysics \& Cosmology (KIPAC), Stanford University, Stanford, CA 94305, USA}\\
{$^{28}$ Department of Astrophysical Sciences, Princeton University, Peyton Hall, Princeton, NJ 08544, USA}\\
{$^{29}$ David A. Dunlap Department of Astronomy \& Astrophysics, University of Toronto, Toronto, ON M5S 3H4, Canada}\\
{$^{30}$ Dunlap Institute for Astronomy \& Astrophysics, University of Toronto, Toronto, ON M5S 3H4, Canada}\\
{$^{31}$ Department of Physics, University of Surrey, Guildford GU2 7XH, UK}\\
{$^{32}$ Department of Astronomy, University of Virginia, Charlottesville, VA 22904, USA}\\
{$^{33}$ Institute for Computational Cosmology, Department of Physics, Durham University, Durham DH1 3LE, UK}\\
{$^{34}$ Lund Observatory, Department of Physics, Lund University, SE\textendash221 00 Lund, Sweden}\\
{$^{35}$ Steward Observatory, University of Arizona, Tucson, AZ 85721, USA}\\
{$^{36}$ University of Colorado Boulder, Boulder, CO 80309, USA}\\
{$^{37}$ Department of Chemistry and Physics, Purdue University Northwest, Hammond, IN 46323, USA}\\
{$^{38}$ Department of Physics, University of Chicago, Chicago, IL 60637, USA}\\
{$^{39}$ Department of Physics, Duke University, Durham, NC 27708, USA}\\
{$^{40}$ Space Telescope Science Institute, 3700 San Martin Drive, Baltimore, MD 21218, USA}\\
{$^{41}$ SLAC National Accelerator Laboratory, Menlo Park, CA 94025, USA}\\
{$^{42}$ Cerro Tololo Inter\textendash American Observatory / NSF NOIRLab, Casilla 603, La Serena, Chile}\\

\end{document}